# Time-resolved measurement of Seebeck effect for superionic metals during a structural phase transition

Shilin Li[1,2,3], Hailiang Xia[4], Takuma Ogasawara[1], Liguo Zhang[1], Katsumi Tanigaki*[1]

[1]*Beijing Academy of Quantum Information Sciences (BAQIS), Bld. #3, No. 10, Xibeiwang East Rd, Haidian District, Beijing, 100193, China*

[2] *Beijing National Laboratory for Condensed Matter Physics, Institute of Physics ( IOP), Chinese Academy of Sciences, 603, Beijing 100190, China*

[3] *University of Chinese Academy of Sciences (UCAS), Beijing 100049, China*

[4]*Beijing Huihaoyu Intelligent Technology Co., Ltd., Room 1604, 1/F, Building 4, Yard 14, Tianhe North Road, Daxing District, Beijing 102699, China.*

Correspondence and requests for materials should be addressed to K.T. (e-mail: katsumitanigaki@baqis.ac.cn).

**ABSTRACT**. We propose a new time (t)-resolved method of both vertical- and horizontal-temperature gradients in an orthogonal configuration (t-resolved T(t)-HVOT) to provide real interpretations of the enhancement in thermoelectric Seebeck effect (SE) observed during the structural phase transition. We apply our new method to superionic-state semiconductors of p-type $Cu_2Se$ and n-type $Ag_2S$. The experimental data differentiate the two types of enhancements during the phase transition: a colossal SE ($S_{\mathrm{colossal}}$), exhibiting an enormous value of 5-10 mV/K, and a slight enhancement in SE ($S_{\mathrm{structure}}$), approximately 1.5-2.0 times larger than those in the absence of the phase transition. We provide critical insights that the enhancement in $S_{\mathrm{colossal}}$ arising during the structural phase transition is not an intrinsic phenomenon. In addition, $S_{\mathrm{structure}}$ also requires more careful discussion, based on the principle of the Seebeck phenomenon.

## I. INTRODUCTION.

The Seebeck effect (SE) is the conversion of temperature difference $\Delta T$ into electric voltage $\Delta V$ at two different locations in a material, discovered by Seebeck in 1821. The thermoelectric phenomena, including the inverse conversion from electric to thermal energy known as the Peltier effect, discovered in 1834, and the Joule-Thomson effect discovered in 1951 in electrical conductors, are currently known as longitudinal thermoelectric phenomena [1–3]. The SE values for typical good thermoelectric materials are generally in the range of 50-500 μV/K, while those for conventional good metals are usually less than 1 μV/K [4–7].

When we reconsider the original physical interpretation of SE, the essential concept is recalled. The coefficient ($S$) of SE, in its theoretical physical sense, is an entropy flow carried out by an elementary charged electron particle upon a given $\Delta T$, as described by $S = -(1/e)(\partial\Omega/\partial N)_E$ in the Kelvin formula [2,8], wherein $\Omega$ is the entropy, e is the elementary charge of an electron, $N$ is the number of elementary particles and $E$ is the internal energy. More careful discussions of such interpretations in the context of strongly correlated electron systems, such as doped Mott-insulators, are reviewed in recent publications [8–10]. A vital consequence of this concept is that further enhancement in SE can be anticipated via co-interplays with other types of entropy. Triggered by this theoretical guideline, numerous intensive studies have been conducted by many scientists to exploit such intriguing possibilities. Such possible entropy terms have included the orbital freedoms of degenerate d- and f-elements, the spin freedom of order/disorder configuration of electron-spin angular momentum in magnetic materials, and the additional thermal entropy of the released/absorbed heat in motion of the constituent elements during a structural phase transition [7,11–15].

The associated phenomena, with the freedoms of orbitals and spins, have been intensively discussed in the past for $Na_xCoO_2$, which exhibits a significantly high S=500μV/K [16,17] even in a good metallic regime. Such experimental data were initially proposed to provide an essential mechanism for the cross-correlated interactions between spin and charge in an entropy flow, leading to a considerable enhancement in SE [17]. The controversial debates, however, have continued to date [18,19]. The spin entropy flow has become another recent active research frontier, as the spin Seebeck effects in the transverse direction [20–22]. The enhancement in SE found during the structural phase transition in the superionic liquid state has also been an intriguing research target [11,12]. An enormous amount of thermal entropy is continuously released as latent heat, and significant movement of the constituent elements takes place in superionic metals during the structural phase transition. Consequently, a substantial enhancement in SE was expected, given its cross-correlated coupling to a charge-flow entropy. Since the degrees of freedom of the constituent elements during a structural phase transition from a low- to high-temperature superionic phase are significantly larger than any other freedoms, such as spin angular momentum [23] and orbitals [24], a vast enhancement in SE was potentially anticipated [13]. Along with this guiding principle, various experimental surveys have been conducted on a superionic conductor $Cu_2Se$, and two types of enhancements have been reported. One is an enhancement in SE ($S_{structure}$), showing a few hundred μV/K, approximately twice the value in the absence of the structural phase transition [12]. Afterwards, an anomalously immense value of SE (named the colossal Seebeck, $S_{colossal}$), reaching several thousands of μV/K, was intriguingly reported under a vertical- and horizontal-gradient ($\nabla_V T$ and $\nabla_H T$) experimental setup [25,26]. Similar anomalous enhancements in SE have also been reported for other superionic semiconductors [27].

The experiments described above garnered considerable attention from other scientists. A super ionic flexible conductor of $Cu_2Se$ showed an anomalously considerable enhancement ($S_{colossal}$) during the structural phase transition, solely when both $\nabla_V T$ and $\nabla_H T$ control are simultaneously provided in the orthogonal configuration (HVOT) [25,26]. The $S_{colossal}$ under the HVOT configuration is significantly greater than the $S_{structure}$ observed solely by $\nabla T_H$ [12]. Sun et al. proposed an energy-band evolution model to explain the enormous SE enhancement in superionic metals based on ARPES measurements [28]. The proposed mechanism's complexity encompasses various origins, including electron-phonon scattering. The real interpretation of the reason for the enhancement in $S$ during the phase transition into superionic metals has remained a subject of controversial scientific debates.

Here, we propose a new time-resolved $T(t)$ modulation method in the orthogonal temperature configuration (time-resolved T(t)-HVOT). In the time-resolved T(t)-HVOT, both steady-state and time-resolved analyses can be performed simultaneously within a single measurement process. Additionally, the structural phase transition can be monitored during the measurements. We study a p-type $Cu_2Se$ and n-type $Ag_2S$ by employing the T(t)-HVOT method. We demonstrate that enhancement in $S_{colossal}$ is not the correct thermoelectric phenomenon. In addition, $S_{structure}$ during the structural phase transition has a physical origin of an increase in resistivity, but delicate arguments should be required based on the fundamental requirements of the Seebeck phenomenon.

## II. METHODS

$Cu_2Se$ and $Ag_2S$ were prepared according to the previous literature [25,29]. We controlled the temperatures at two positions on a specimen, 10 mm long and 1mm$^2$ in cross-sectional area, as a function of time ($t$) with an accurate increase rate of $T(t)$ as small as $6.8\times10^{-4}$ K/sec in the vertical direction using a large-capacity bottom heater. The $T(t)$ at one side was further modulated in the horizontal direction from low to high $T$ at a higher rate of $6.7\times10^{-1}$ K/sec repeatedly over many cycles by varying the input current intensity of a small-capacity heater. In this measurement process, a time-resolved $T(t)$ difference of $\Delta T(t)$ was generated with high accuracy, so that $\Delta T(t)$ repeatedly can change its sign from minus to plus as shown in Fig.1(a, b). The generated thermoelectric voltage $\Delta V(t)$ was simultaneously monitored as a function of $t$. Using the collected data sets, time-resolved SE coefficients of $S(t)_{\text{time-T(t)-HVOT}}$ were evaluated in the limit of the infinitesimal time resolution of the measurements. Finally, the $S(t)$ values were converted to $S[T(t)]$-$T(t)$ relations, as described in supplementary SFig.1 and SFig.2.

## III. RESULTS

### Time(t)-resolved T(t)-HVOT configuration

The time-resolved T(t)-HVOT configuration is shown in Fig.1(a). First, we describe the critical information available from this new measurement technique for evaluating SE coefficients during a structural phase transition. A small capacity of the second heater provides a well-controlled time-modulated $T_2(t)$ in the horizontal (H)-direction of $T$-gradient $\nabla_H T_2(t)$ with a relatively fast rate of 0.1K/sec (25 measurement points, dT=0.0687 K, and dt=0.687 sec in the time resolution) in each modulation process. On the other hand, $T_1(t)$ is dominantly controlled by a large-capacity base heater with a tiny gradient $\nabla_V T_1(t)$ of 0.0216 K/sec in the vertical direction. During this $T(t)$-HVOT periodic control process, we additionally set a $t$=500 sec as a waiting time to generate a constant temperature at the end of every $T_2(t)$ modulation process, so that conventional steady state measurements with a time independent $\Delta T$ can be additionally given in the same measurement process, as shown in Fig.1(a,b).

In the time-resolved T(t)-HVOT measurements, the evolution of the structural phase transition can be sensitively monitored by a $T$=$T_1(t)$-$T_2(t)$ response as a function of $t$. In contrast, $T_1(t)$ increased linearly at a very slow rate, showing minor sensitivity to the structural phase transition. In these well-controlled $T$ regulations, not only multiple cycles of time-resolved $\Delta T(t)$ but also a conventional steady-state constant $\Delta T$, before starting each modulation cycle, were repeated. The $T_2(t)$ evolution exhibited a dip, as shown in Fig.1(d), indicating heat absorption during the structural phase transition. We simultaneously observed the generated thermoelectric voltage $\Delta V(t)=V_2(t)-V_1(t)$, corresponding to $\Delta T(t)=T_2(t)-T_1(t)$, at an identical time $t$ between two positions $x_1(T(t))$ and $x_2(T(t))$ of a sample. A voltage with a sign of $\pm\Delta V(t)$ is observed as the thermoelectric response to $\pm(\mp)\Delta T(t)$, corresponding to p(n)-type carriers of materials. Three types of analyses were provided. We first measured thermoelectric $\Delta V(t)$ as a function of time, and the SE coefficients were evaluated, based on the general definition of $S(t) = -\Delta V(t)/\Delta T(t)$. Such an evaluated $S(t)$ value is denoted as $S(t)_{\text{steady-T(t)-HVOT}}$. As the second evaluation, by taking the time evolution taking into consideration, the $S(t)$ values were evaluated by $S(t)_{\text{time-T(t)-HVOT}} = -\text{d}[\Delta V(t)/\text{d}t]/\text{d}[\Delta T(t)/\text{d}t]$ in the infinitesimal time of dt within the time resolution of our experiments, as shown by the red data in Fig.1(b,c). As the third type of analysis, a conventional steady-state SE coefficient $S_{\text{c-steady}}$ was evaluated by collecting experimental points under steady-state equilibrium with $\Delta T$ constant. The details of these three types of analyses are described in the supplementary SFig.1 and SFig.2.

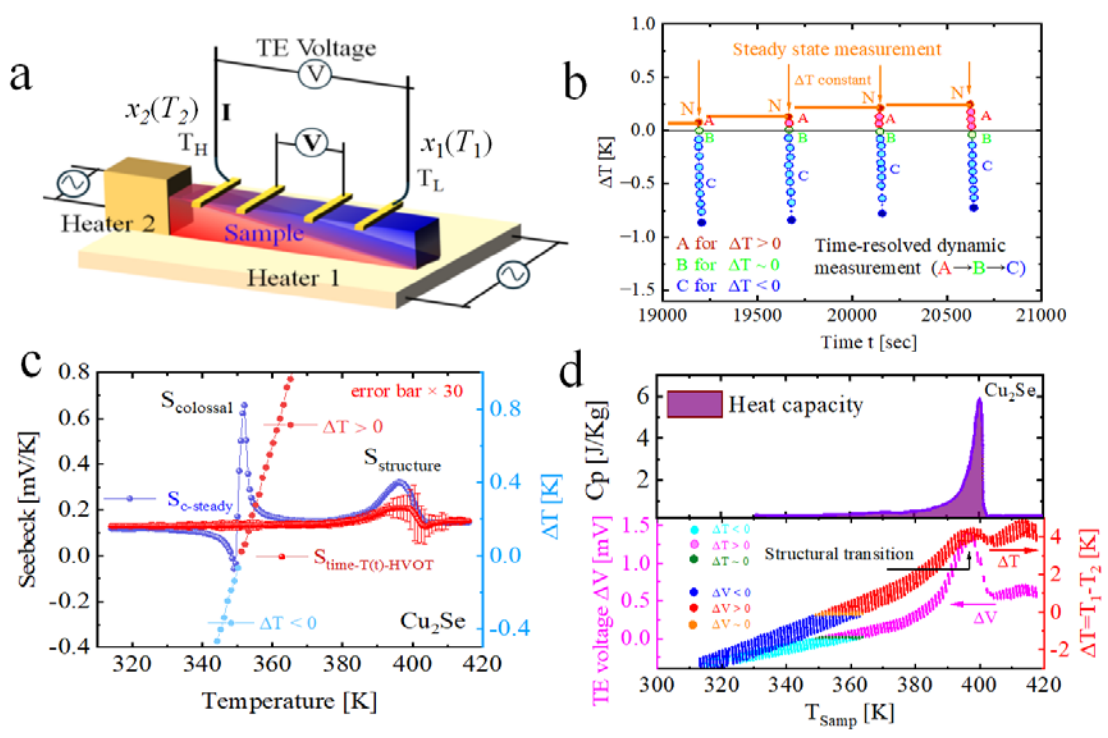

**Fig.1** (a) Time($t$)-resolved T(t)-HVOT with both horizontal and vertical $T$ control. The $T$ gradient in the horizontal direction of $\Delta_H T(t) = T_1(t)-T_2(t)$ is modulated as a function of time at a faster rate than $\Delta_V T(t)$ controlled by the bottom heater. (b) $\Delta T(t)$ as a function of $t$. Before starting a new modulation sequence of $T_2(t)$, a reasonably long waiting period of 500 sec was provided for each modulation process, to enable a conventional steady-state analysis ($S_{\text{c-steady-HVOT}}$). The time-resolved T(t)-HVOT data sets provide other two types of SE analyses of $S_{\text{time-T(t)-HVOT}}$ and $S_{\text{steady-T(t)-HVOT}}$, in addition to the $S_{\text{c-steady-HVOT}}$. The analytical details are described in SFig.1. (c) Two types of thermoelectric SE coefficients, $S_{\text{colossal}}$ and $S_{\text{structure}}$, were observed for $Cu_2Se$ in $S_{\text{c-steady-HVOT}}$. In contrast, $S_{\text{colossal}}$ smeared out in the $S_{\text{time-T(t)-HVOT}}$ analysis. The error bars are indicated. $S_{\text{colossal}}$ appears with a sign change at $\Delta T(t) \approx 0$. (d) $\Delta T(t)$ and $\Delta V(t)$ were measured as a function of $t$. The heat capacity $C_p$ is shown together by correcting its position to that at the surface of a sample in the time-resolved T(t)-HVOT. We measured $S(T(t))$ as a function of $T(t)$ for both p-type $Cu_2Se$ (phase transition enthalpy of $\Delta H_r = 32$ J/g [13]) and n-type $Ag_2S$ ($\Delta H_r = 15$ J/g [29–31]). The resulting thermoelectric data are explained in the next section, together with resistivities measured under the same temperature modulation sequences.

### Thermoelectric data of $Cu_2Se$ by time-resolved T(t)-HVOT

Fig.1c,d shows $S_{\text{c-steady}}$ and $S_{\text{time-T(t)-HVOT}}$ for $Cu_2Se$ with p-carriers obtained from the analyses of the t-resolved T(t)-HVOT measurements. In the $S_{\text{c-steady}}$, a large $S_{\text{colossal}}$ [25,26] (displayed by the blue line) was analyzed at a $T$ away from the phase transition temperature, together with a small enhancement in $S_{\text{structure}}$ [13,32] (displayed as the same blue line) as shown in Fig.1c. The latter was observed at a similar temperature to that of the structural phase transition. As justified by more careful examinations described later, $S_{\text{colossal}}$ is observed when $\Delta V(t)$ is finite but not zero under the condition that the zero-denominator term $\Delta T(t) = 0$, hitherto leading to a vast $S_{\text{colossal}}$ that is divergently enormous. It is essential to note that the $S_{\text{colossal}}$ signal smeared out when we employed the $t$-resolved analysis of $S_{\text{time-T(t)-VHOT}}$, leaving only $S_{\text{structure}}$ (red line) with its weakened intensity, as shown clearly in Fig.1c.

In the $S_{\text{colossal}}$, both positive and negative signs were observed as previously reported [25,26]. As clearly understood by the polarity of $\Delta T(t)$ in Fig.1c, the reason for the sign change in $S_{\text{colossal}}$ is due to the crossing into the opposite polarity of $\Delta T(t)$ and a generated finite thermoelectric $\Delta V(t)$ with the same sign. Although a local equilibrium, in a small area of the material, is achieved, a global system equilibrium, in the whole size of the material, is hardly attained during the structural phase transition, as described later in the discussion. The situation is strongly dependent on the experimental condition (see supplementary SFig.3, measured using a different time sequence). The experimental evidence so far explained unambiguously indicates the fact that the origin of the colossal SE is $S_{\text{colossal}} = -\Delta V(t,\ x_{\neq}x_1,x_2)/(T(t,\ x_1)-T(t,\ x_2))$, which does not arise from the identical locations of $x_1(t)$ and $x_2(t)$ where the temperatures were monitored, at the same measurement time $t$. The requirements for the intrinsic SE definition of both identical time $t$ and position $x(t)$ are violated as described in the discussion. This is also why the immense value of $S_{\text{colossal}}$ is observed at a position away from the phase transition temperature with a sign change.

### Thermoelectric Seebeck values of $Ag_2S$ by time-resolved T(t)-HVOT

Similar experiments were performed for $Ag_2S$ with electron carriers using the time-resolved T($t$)-HVOT (dt = 0.036 sec, dT = 0.036 K, and $\Delta$t/cycle = 0.9 sec/cycle). A similar dip in $T_2(t)$ to that of $Cu_2Se$, reflecting the latent heat absorption [29] at T = 435 ~ 445 K during the structural phase transition, was observed (Fig.2a). By reflecting the n-type carriers of $Ag_2S$, which differ from p-type $Cu_2Se$, the majority of the experimental data were a response of $\pm\Delta T(t)/\mp\Delta V(t)$.

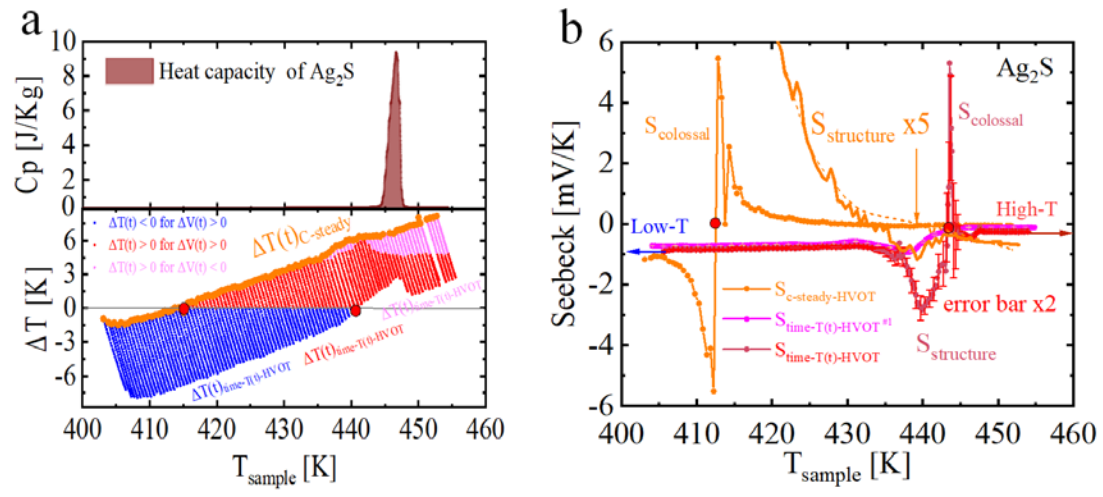

**Fig.2** Thermoelectric SE data of $Ag_2S$ together with its heat capacity ($C_p$). (a) Temperature difference $\Delta T(t)=T_1(t)-T_2(t)$. A dip in $\Delta T(t)$ was observed, corresponding to the latent heat during the structural phase transition. (b) Evaluated SE values as a function of $T_{sample}(t)$ (= $(T_1(t)+T_2(t))/2$). Both anomalously huge $S_{colossal}$ and small $S_{structure}$ (orange line) were observed in the $S_{c\text{-}steady}$ (orange line) analysis. However, $S_{colossal}$ smeared out in the $S_{time\text{-}T(t)\text{-}HVOT}$ (red line) analysis with solely remaining a $S_{structure}$ with a large decrease in its intensity. The error bars are displayed. After the corrections of the $S_{colossal}$ influence, the intensity of $S_{structure}$ (pink line) was further reduced. The analytical details are described in SFig.2.

Although the measurement sequence of time-resolved T($t$)-HVOT was the same as that for $Cu_2Se$, the collected data of $\Delta T(t)$ and $\Delta V(t)$ of $Ag_2S$ as a function of $t$ were very dissimilar from those of $Cu_2Se$. The significant difference is that an indelibly large non-zero thermoelectric voltage $\Delta V(t)$ is generated in a wide $\Delta T(t)$ in the vicinity of the $\Delta T(t)$=0 zone (see, SFig.2). This clearly indicates that we are not guaranteed to monitor the intrinsic thermoelectric voltage generated by the temperature difference at the identical time $t$ and place $x(t)$ as described earlier. The enhancements observed for anomalous $S_{colossal}$ by the two-types steady state analyses of $S_{steady\text{-}T(t)\text{-}HVOT}$ and $S_{c\text{-}steady}$ are very huge of 6-10 mV/K, as shown in SFig.2. The two anomalous $S_{colossal}$ signals were analyzed corresponding to the position of $\Delta T(t) \cong 0$, as explained for $Cu_2Se$. The $S_{colossal}$ signal located away from the structural phase transition disappeared when the analysis was undertaken using $S_{time\text{-}T(t)\text{-}HVOT}$, leaving a $S_{structure}$. These are consistent with the situation of $Cu_2Se$. It is noted that the other $S_{colossal}$ observed $\Delta T(t) \cong 0$ during the structural phase transition was still left in the $S_{time\text{-}T(t)\text{-}HVOT}$ due to the large latent heat, not to be ignored in the case of the inhomogeneous $Ag_2S$ sample. After the corrections of the $S_{colossal}$ influence, significantly reduced intensity of $S_{structure}$ was evaluated (the pink line, see more details in supplementary SFig.2).

## IV. DISCUSSION

Two possible reasons are considered for the SE coefficients increasing to an enormous mV/K level. One is the constituent-element motion under a super-ionic state by greatly modulating a carrier number as suggested previously [25], and the other is the phonon drag [33–37]. However, our experimental data rule out the former because we did not observe such a change in electrical resistivity ρ that indicates an extensive carrier modulation of the opposite sign. In addition, the phonon drag effect requires low temperatures and is generally accompanied by an increase in thermal conductivity [38], which is also not evident in the present case.

When we come back to the original definition of SE coefficient, $S(T) = -[\frac{1}{T1-T2}] \int_{x1(T1)}^{x2(T2)} (dV(x))/dT(x))(dT(x)/dx)dx$ ], a local equilibrium of a segment i of $S_i[T_i(x_i(t))] = -dV_i[x_i(t)]/dT_i[x_i(t)]$ at an observation time of $t$ is continuously connected to a global system equilibrium: $S(t\rightarrow\infty) = -(V_1(x_1)-V_2(x_2))/(T_1(x_1)-T_2(x_2))$ between $x_1(t)$ and $x_2(t)$ as t increases, where the equilibrium is simultaneously realized over the whole measurement segments. In the definition of SE coefficient, the system global equilibrium between $x_1(T_1(t))$ and $x_2(T_2(t))$ is required, where $x_1(T_1(t))$ and $x_2(T_2(t))$ are the two positional ends of a material having temperatures of $T_1(t)$ and $T_2(t)$, respectively. For evaluating intrinsic $S(T)$, a global equilibrium state is assumed to be realized in conventional materials, given a sufficient waiting time to reach it. However, this is not the case for a structural phase transition. Considering the non-negligible $T(x(t))$ fluctuation during a structural phase transition, a latent heat is generated or absorbed in various positions of a sample specimen at around the phase transition temperature. In the conventional steady-state measurements, however, temperatures of the two endpoints of a sample, $T(x_1(t))$ and $T(x_2(t))$, are used for SE evaluations by assuming the system's global equilibrium. In contrast, the thermoelectric voltage $\Delta V(t) = \lim_{\Delta x(t)\rightarrow 0} \sum_i \Delta V_i[\Delta x(t)] = \int_{x1}^{x2} V(x(t))dx(t)$ is an integration of each segment at a time of $t$, and the time for reaching the electronic-state equilibrium is greatly shorter than that for the thermal equilibrium. As a result, the global $\Delta V(t)$ can have a non-zero value despite $\Delta T(t) = 0$ by violating a global equilibrium requirement of $\Delta V_i[\Delta x(t)]=0$ for $\Delta T_i[\Delta x(t)]=0$, even if a local equilibrium is satisfied. Such a situation is unambiguously evident in the case of $Ag_2S$, as explained earlier, where non-zero $\Delta V(t)$ appears for $\Delta T(t) = 0$ over a wide range of $t$.

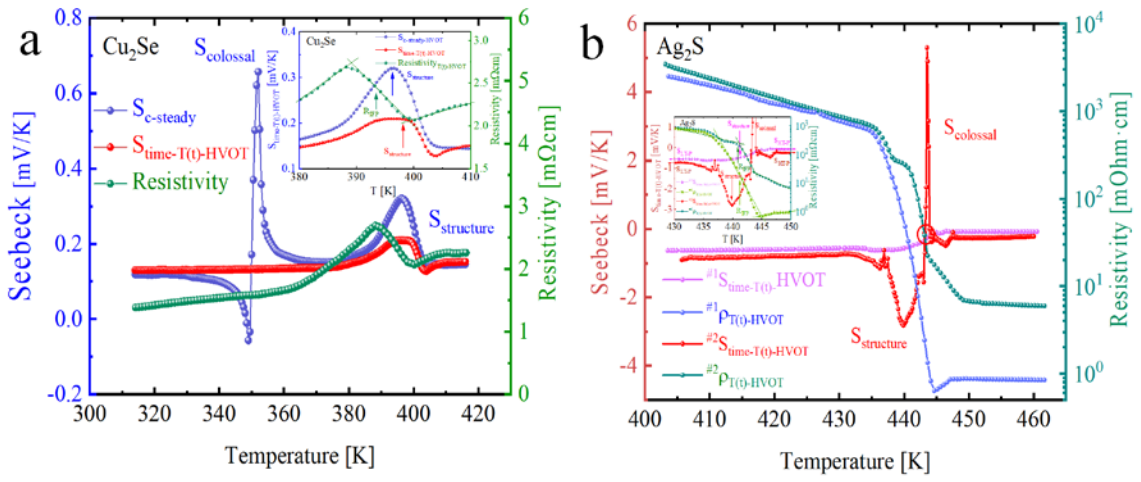

**Fig.3** (a) SE value ($S$) and resistivity (ρ) of $Cu_2Se$ with p-type carriers analyzed by $S_{\text{c-steady}}$ and $S_{\text{time-T(t)-HVOT}}$. The observed $S$ is reported as the average $S$ over each $\Delta T$(t) modulation sequence (the details are described in SFig.1). An enhancement in $S_{\text{structure}}$ is observed by $S_{\text{c-steday}}$ and $S_{\text{time-T(t)-HVOT}}$, and their enhancements occur when the ρ increases during the phase transition. (b) $S$ and ρ of $Ag_2S$ with n-type carriers evaluated by $S_{\text{c-steday}}$ and $S_{\text{time-T(t)-HVOT}}$ analyses based on the T(t)-HVOT data. (see the details in SFig.2). Reflecting the n-type carriers, $S$ with a negative sign was observed in the absence of a structural phase transition. The inset figures of Fig.3(a, b) show a discrepancy between the maximum of $|S_{\text{structure}}|$ and the infection point of ρ ($R_{\text{IFP}}$), which is critical for further careful arguments on $S_{\text{structure}}$, as described in the text.

According to the SE coefficient in the framework of the Boltzmann transport equation, $S(T) = \int_{BZ} dk\, (1/T)[((-e/3)D(\varepsilon(k))\tau(\varepsilon(k))v(\varepsilon(k))^2(\varepsilon(k)-\mu)(\partial f/\partial\varepsilon(k)) /[\int_{BZ} dk\, D(\varepsilon(k))\tau(\varepsilon(k))v(\varepsilon(k))^2(\partial f_{FD}/\partial\varepsilon(k))]$, where $f_{FD}$ is the Fermi-Dirac distribution function, $D(\varepsilon(k))$ is the density of states, $\tau(\varepsilon(k))$ is the relaxation time, $v(\varepsilon(k))$ is the group velocity, and the integration is made over the entire 1st Brillouin zone (BZ). Since the denominator in the equation of $S(T(t))$ corresponds to the electrical conductivity σ (=1/ρ) dominated by the carrier density at the Fermi level, an increase in resistivity gives an enhancement in $S(T(t))$. The $S$ values evaluated in our time-resolved T(t)-HVOT showed a higher $S$ = 130 μV/K with a higher ρ of $4.6\text{x}10^{-1}$ Ω·cm in the low-T phase than $S$=124 μV/K of the high-T phase with the lower ρ of $4.3\text{x}10^{-2}$ Ω·cm, and these are consistent with the general theory. A similar correspondence, between the enhancement in $S_{\text{structure}}$ and an increase in $\rho$ during the structural phase transition, is also reported in the previous literature [11,12]. Our T(t)-HVOT showed that $S_{\text{structure}}$ is enhanced because of the increase in ρ via the grain boundary scattering during the structural phase transition. However, the peak positions of the $S_{\text{structure}}$ and the inflection point of the ρ ($R_{\text{IFP}}$) have a discrepancy beyond the experimental errors for both $Cu_2Se$ and $Ag_2S$, as can be displayed in the inset figures of Fig.3a and b. Furthermore, the strength of $S_{\text{structure}}$ changes sensitively upon measurements (see the inset of Fig.3b). These experimental facts indicate unambiguously that the essential requirement of the identical time and location for the SE definition is also violated even in the case of $S_{\text{structure}}$. The real discussion on $S_{\text{structure}}$ should be made more carefully. There is no experimental evidence, so far, about the contribution of cross-conversion from other entropies during the structural phase transition.

The $t$-resolved T(t)-HVOT method unambiguously exhibited that the peak position between S and ρ are not at the identical time and place as seen evidently in Fig.3. This suggests that although the scattering among the grain boundaries is the primary reason for the $S_{\text{struture}}$ [9], the requirement for the intrinsic SE phenomena is not fully fulfilled within the accuracy of experiments even in the case of $S_{\text{structure}}$. According to the microscopic origin of SE in its physical meaning is the electric field of $\widetilde{E} = E_0-(1/(-e))(d\mu(T(x))/dx)(dx/dT(x))$ generated by the $\nabla T$ [39], where $\mu(T(x))$ is the chemical potential at x(T(t),t), the identical time and place between $\nabla E$(x,t) and $\nabla T$(x,t) is essential. Consequently, the enhancement in $S_{\text{structure}}$ is also most likely extrinsic, or it should be significantly smaller than what we observed within the experimental accuracy.

## V. CONCLUSIONS

Ever since the discovery of a large SE coefficient S for $Na_xCoO_2$ in its metallic states, a critical question has continued about whether SE can be enhanced via the cross-conversion of other entropies, such as the freedom of orbitals and spin angular momentum. More recently, various intensive studies have been conducted on superionic metals, exhibiting a structural phase transition. Two types of enhancements, $S_{\text{structure}}$ and $S_{\text{colossal}}$ were reported. We proposed a new time-resolved T(t)-HVOT measurement method as a powerful tool for obtaining the intrinsic interpretations. The time-resolved T(t)-HVOT provided both steady-state and time-resolved information during a single measurement process, together with the information of the structural phase transition. We showed that $S_{\text{colossal}}$ is not an intrinsic SE phenomenon. We concluded that the enhancement in $S_{\text{structure}}$ is not due to entropy conversion from the latent heat entropy during the structural transition. The $S_{\text{structure}}$ reported in the past is greatly overestimated or nearly zero, because the requirements of identical time and position based on the definition of SE phenomena are violated in the steady-state measurements.

## ACKNOWLEDGMENTS

Measurements of electrical conductivity were conducted at the Nano-platform facility, Beijing Academy of Quantum Information Sciences (BAQIS), China, and the Advanced Institute of Materials Science (AIMR), Tohoku University, Japan. We acknowledge Tsunehiro Takeuchi for supplying high quality $Cu_2Se$ and $Ag_2S$. KT thanks Yinshang Liu for his technical support in developing the temperature control sequence program using the Labview program code. This research project was supported by Program Innovation for Quantum Science and Technology (Gr ant No. 2023ZD0300500), the National Natural Science Foundation of China (NSFC Grant No. 12174027), Beijing Natural Science Foundation (Grant No. IS25007), and the CREST project by JST on Thermal Management.

# Supplementary of Time-resolved measurement of Seebeck effect for superionic metals during structural phase transition

Shilin Li[1,2,3], Hailiang Xia[4], Takuma Ogasawara[1], Liguo Zhang[1], Katsumi Tanigaki*[1]

[1]*Beijing Academy of Quantum Information Sciences (BAQIS), Bld. #3, No. 10, Xibeiwang East Rd, Haidian District, Beijing, 100193, China*

[2] *Beijing National Laboratory for Condensed Matter Physics, Institute of Physics ( IOP), Chinese Academy of Sciences, 603, Beijing 100190, China*

[3] *University of Chinese Academy of Sciences (UCAS), Beijing 100049, China*

[4]*Beijing Huihaoyu Intelligent Technology Co., Ltd., Room 1604, 1/F, Building 4, Yard 14, Tianhe North Road, Daxing District, Beijing 102699, China.*

Correspondence and requests for materials should be addressed to K.T. (e-mail: katsumitanigaki@baqis.ac.cn).

**A. Thermoelectric data of $Cu_2Se$ observed by time-resolved T(t)-HVOT.**

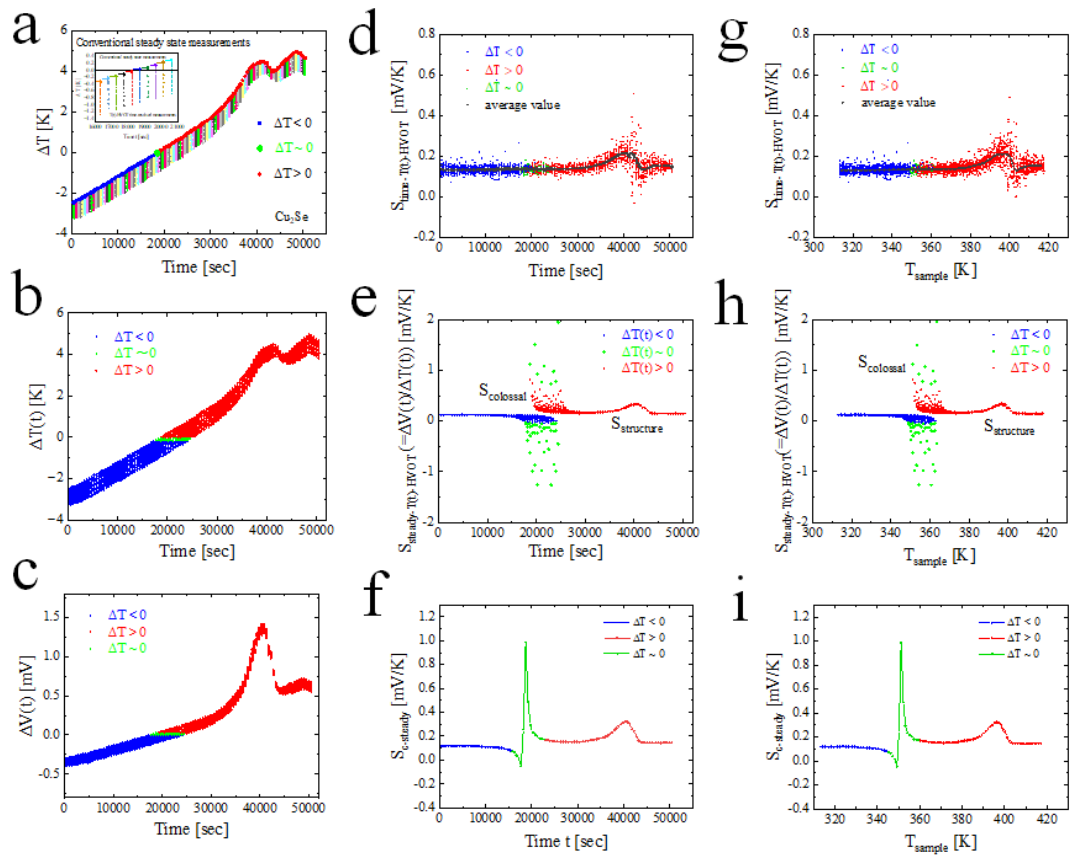

**SFig1**. (a) Time-resolved temperature sequences (different-color lines). We use a large-capacity heater to control the overall temperature of the sample, and a small-capacity second heater to periodically heat a single position (Heater 2 in Fig.1a) of the sample. A 500-sec waiting time was provided between cycles to stabilize the temperature gradient (the temperature error was smaller than 0.01 K). The first data point of each cycle (called the time $t_o$) can be used for the conventional steady state analysis of $S_{c\text{-}steady}$ (b) Modulated temperature difference $\Delta T(t)=T_1(t)-T_2(t)$. (c) Thermoelectric voltage $\Delta V(t)$ measured at time $t$. (d-f) SE values were evaluated by the three analytical methods, such as $S_{time\text{-}T(t)\text{-}HVOT}(t) = -d[(\Delta V(t))/dt]/d[(\Delta T(t)/dt)]$, $S_{steady\text{-}T(t)\text{-}HVOT}(t) = -(\Delta V(t))/(\Delta T(t))$, and $S_{c\text{-}steady}(t) = -(\Delta V(t_0))/(\Delta T(t_0))$, as a function of time $t$. (g-i) $S_{time\text{-}T(t)\text{-}HVOT}(T)$, $S_{steady\text{-}T(t)\text{-}HVOT}(T)$, and $S_{c\text{-}steady}(T)$ as a function of $T_{sample}(t) =(T_1(t)+T_2(t))/2$. The $S_{steady\text{-}T(t)\text{-}HVOT}$ analytical method directly shows the relationship between $\Delta T(t)$ and $\Delta V(t)$ as a function of time $t$, as shown SFig.1e and h. $S_{c\text{-}steady}$ and $S_{time\text{-}T(t)\text{-}HVOT}$ analyses show both $S_{colossal}$ and $S_{structure}$ due to the lack of the global equilibrium, as described in the main text. However, $S_{colossal}$ disappeared in the $S_{time\text{-}T(t)\text{-}HVOT}$ analysis, and $S_{structure}$ also became smaller.

**B. Thermoelectric data of n-type $Ag_2S$ The time-resolved T(t)-HVOT.**

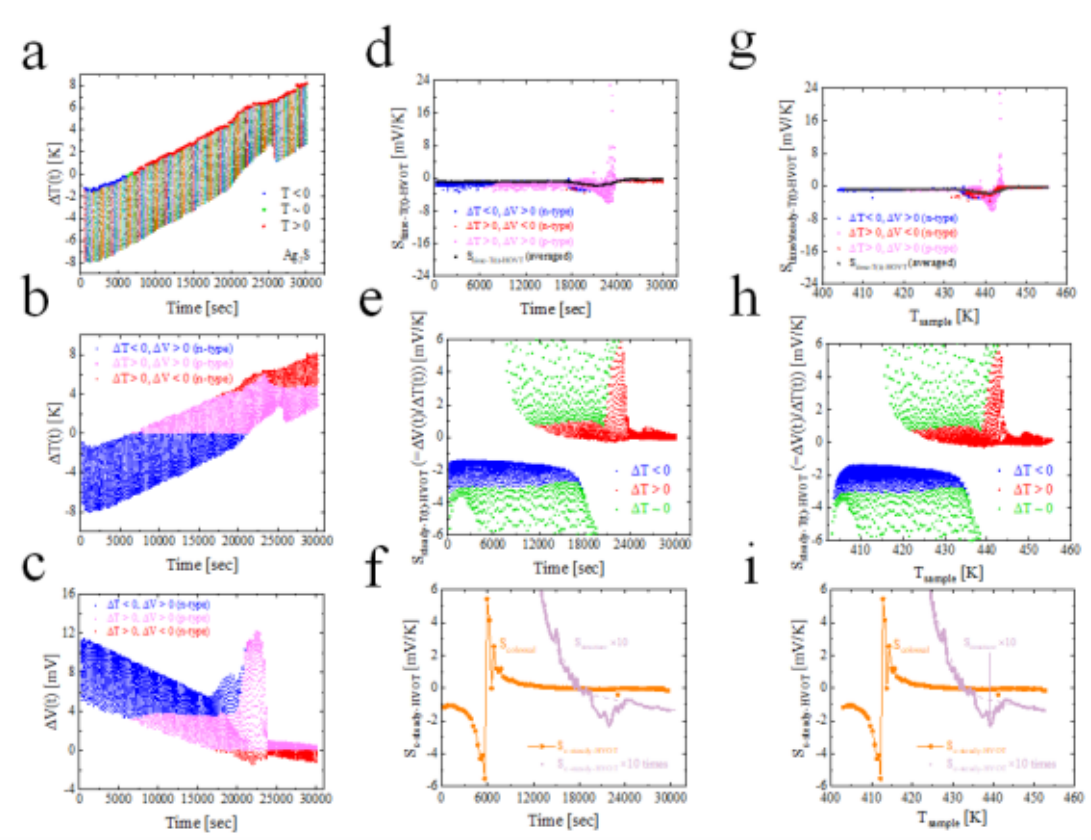

**SFig.2** (a) Time-resolved temperature sequences (different-color lines), including the conventional steady-state sequence. (b) Temperature difference $\Delta T(t)=T_1(t)-T_2(t)$. (c) Generated thermoelectric voltage $\Delta V(t)$ corresponding to $\Delta T(t)$. (d-f) $S_{time\text{-}T(t)\text{-}HVOT}(t) = -[d(\Delta V(t))/dt]/[d(\Delta T(t))/dt]$, $S_{steady\text{-}T(t)\text{-}HVOT}(t) = -(\Delta V(t))/(\Delta T(t))$ and $S_{c\text{-}steady}(t) = -(\Delta V(t_0))/(\Delta T(t_0))$ as a function of time $t$. (g-i) $S_{time\text{-}T(t)\text{-}HVOT}(T) = -d(\Delta V(t))/d(\Delta T(t))$, $S_{steady\text{-}T(t)\text{-}HVOT}(T) = -(\Delta V(t))/(\Delta T(t))$ and $S_{c\text{-}steady}(T) = -(\Delta V(t_0))/(\Delta T(t_0))$ as a function of $T_{sample}$. Since $\Delta T(t)\simeq 0$ overlaps the structural phase transition and $Ag_2S$ is more inhomogeneous than $Cu_2Se$, a signal of $S_{colossal}$ remained even in the $S_{time\text{-}}$ $_{T(t)\text{-}HVOT}$(T) method as shown in SFig.2( d, g). After making corrections of the influences of $S_{colossal}$ signal, the real $S_{structure}$ was evaluated to be greatly smaller. Because the maximum of $|S_{structure}|$ and the reflection point of resistivity ρ show a discrepancy in position, the intrinsic understanding of $S_{structure}$ needs further careful consideration, as described in the main text.

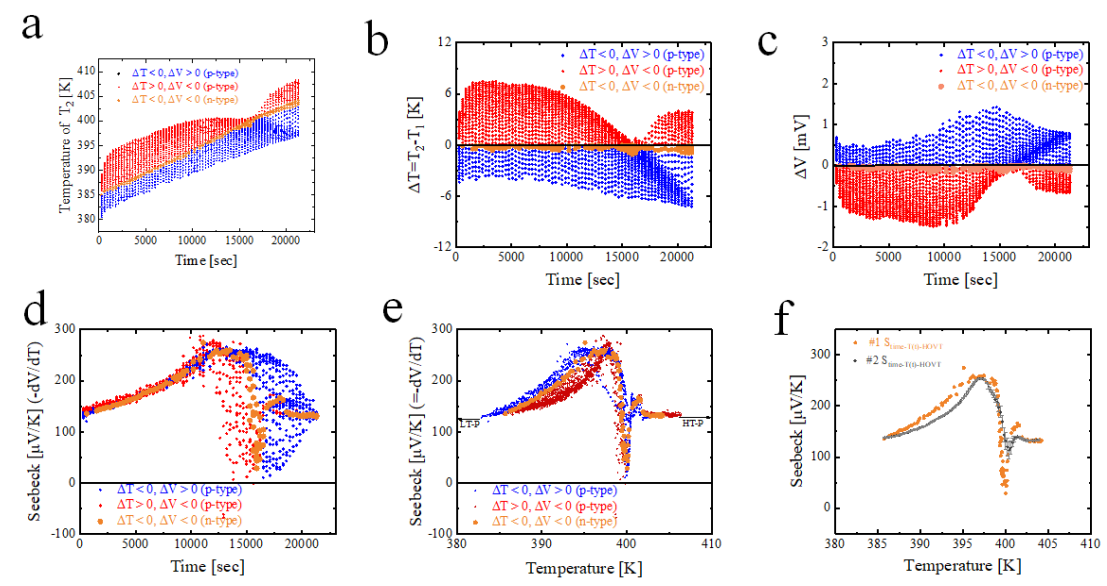

**SFig.3 A different time sequence in time-resolved T(t)-HVOT measurements for $Cu_2Se$.** (a) The modulation of temperature $T_2$(t). (b) Temperature difference $\Delta T(t)=T_1(t)\text{-}T_2(t)$. (c) Generated thermoelectric voltage $\Delta V(t)$ corresponding to $\Delta T(t)$. (d-f) $S_{time\text{-}HVOT}(t)$, $S_{time\text{-}HVOT}(T)$ and the average of $S_{time\text{-}HVOT}(T)$. The measurements during the phase transition depend on the time-resolved sequence. A different $\Delta T(t)$, with a relatively higher increase than that described in the text and SFig.1, was used for $Cu_2Se$ with a rate of 0.67 K/sec and $T_2$(t) modulation in the time-resolved T(t)-HVOT method. As long as the modulation temperature rate is sufficiently sensitive to detect the structural phase transition, a dip corresponding to it was observed. However, the ambiguity of the collected thermoelectric voltage data at $\Delta T(t)\simeq 0$ greatly increased. For obtaining the physically meaningful experimental data during the structural phase transition, a sufficiently low rate of increase and highly time-resolved modulation are critical.

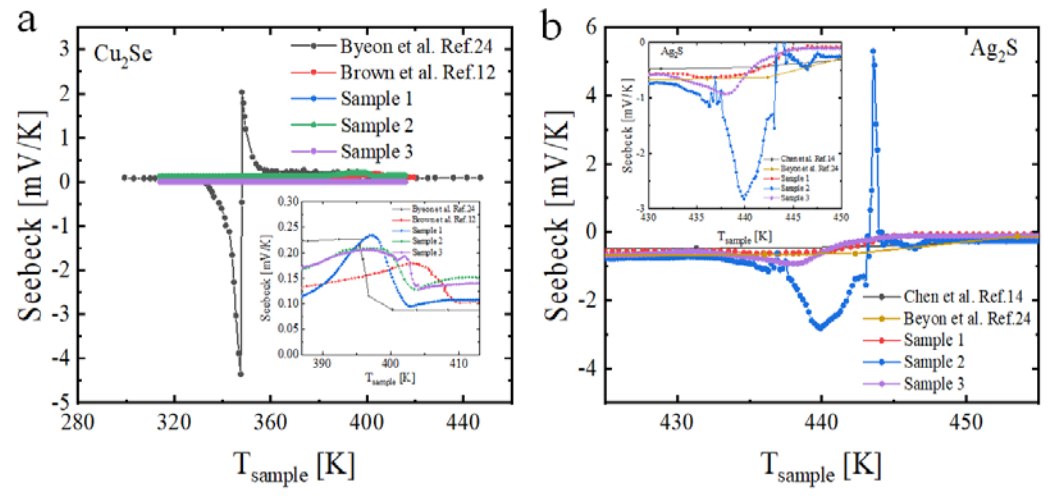

**SFig.4** Comparison of $S_{colossal}$ and $S_{structure}$ between our present work and those of the previous reports for $Cu_2Se$ (a) and $Ag_2S$ (b). The phase transition temperature, $S_{structure}$, and $S_{colossal}$ are experimentally reproduced. However, the intrinsic interpretations need careful consideration, based on the identical time and space of the fundamental SE phenomenon as described in the main text.